\documentclass[sigconf]{acmart}
\usepackage{enumitem}
\usepackage{multirow}
\usepackage{graphicx}
\usepackage{booktabs}
\usepackage{verbatim}
\usepackage{float}
\usepackage{threeparttable}
\usepackage{balance}
\PassOptionsToPackage{type={none}}{doclicense}
\AtBeginDocument{%
  }

\begin{document}

\title{A Soft-partitioned Semi-supervised Collaborative Transfer Learning Approach for Multi-Domain Recommendation}

\author{Xiaoyu Liu}
\email{liuxiaoyv@buaa.edu.cn}
\affiliation{%
  \department{Institute of Artificial Intelligence}
  \institution{Beihang University}
  \city{Beijing}
  \country{China}
}

\author{Yiqing Wu}
\email{wuyiqing20s@ict.ac.cn}
\affiliation{%
  \department{Institute of Computing Technology}
  \institution{Chinese Academy of Science}
  \city{Beijing}
  \country{China}
}

\author{Ruidong Han}
\email{hanruidong@meituan.com}
\affiliation{%
  \institution{Meituan}
  \city{Beijing}
  \country{China}
}

\author{Fuzhen Zhuang}
\authornote{Corresponding author.}
\authornote{Fuzhen Zhuang is also at State Key Laboratory of Complex \& Critical Software Environment, Beijing, China.}
\email{zhuangfuzhen@buaa.edu.cn}
\affiliation{%
  \department{Institute of Artificial Intelligence}
  \institution{Beihang University}
  \city{Beijing}
  \country{China}
}

\author{Xiang Li}
\email{lixiang245@meituan.com}
\affiliation{%
  \institution{Meituan}
  \city{Beijing}
  \country{China}
}

\author{Wei Lin}
\email{linwei31@meituan.com}
\affiliation{%
  \institution{Meituan}
  \city{Beijing}
  \country{China}
}

\renewcommand{\shortauthors}{Xiaoyu Liu et al.}

\begin{abstract}
  In industrial practice, Multi-domain Recommendation (MDR) plays a crucial role. Shared-specific architectures are widely used in industrial solutions to capture shared and unique attributes via shared and specific parameters. However, with imbalanced data across different domains, these models face two key issues: (1) \textbf{Overwhelming}: Dominant domain data skews model performance, neglecting non-dominant domains. (2) \textbf{Overfitting}: Sparse data in non-dominant domains leads to overfitting in specific parameters. To tackle these challenges, we propose \textbf{S}oft-partitioned \textbf{S}emi-supervised \textbf{C}ollaborative \textbf{T}ransfer \textbf{L}earning (\textbf{SSCTL}) for multi-domain recommendation. SSCTL generates dynamic parameters to address the overwhelming issue, thus shifting focus towards samples from non-dominant domains. To combat overfitting, it leverages pseudo-labels with weights from dominant domain instances to enhance non-dominant domain data. We conduct comprehensive experiments, both online and offline, to validate the efficacy of our proposed method. Online tests yielded significant improvements across various domains, with increases in GMV ranging from 0.54\% to 2.90\% and enhancements in CTR ranging from 0.22\% to 1.69\%. 
\end{abstract}

\ccsdesc[500]{Information systems~Recommender systems}

\keywords{Recommender System; Semi-supervised Learning; Multi-Domain Recommendation}
\maketitle

\section{Introduction}
Large-scale commercial platforms often span multiple domains. A typical e-commerce platform homepage includes a primary recommendation list and multiple sub-domains. However, existing multi-domain recommendation methods treat all domains equally, overlooking the fact that the main feed often dominates traffic, leading to highly imbalanced data—a challenge that remains largely unaddressed.
As shown in Table \ref{tab:overlap}, data from one week of operations reveal that the homepage dominates, accounting for over 80\% of traffic, while some domains contribute less than 1\%. Traditional single-domain models face two key challenges: (1) They fail to leverage cross-domain data and transferable knowledge, despite significant user and item overlap. (2) Data imbalance creates sparsity issues, leading to suboptimal performance.

Significant efforts have been undertaken \cite{ma2018modeling,tang2020progressive,chang2023pepnet,yang2022adasparse,luo2025aread,luo2025cdc} have been made to tackle \textbf{M}ulti-\textbf{D}omain \textbf{R}ecommendation (MDR) problems. Recently, shared-specific parameter architectures, such as HMoE\cite{li2020improving} and STAR\cite{STAR}, have proven effective by capturing domain-specific attributes with specific parameters and cross-domain commonalities with shared parameters.
However, these methods typically use a \textbf{hard-partitioning} approach , relying solely on domain indicators to divide data. This approach struggles with uneven data distribution, leading to two significant issues:

\begin{itemize}[leftmargin=*]
    \item The sparsity of non-dominant domains poses a challenge to the learning of specific parameters, often resulting in overfitting.
    \item For shared parameters, severe data imbalance can cause the model to be dominated by the dominant domain's data \cite{yang2022adasparse,chang2023pepnet}. While some studies address this by using domain indicators to distinguish instances, the imbalance may lead the model to overlook these indicators, exacerbating the dominance issue \cite{dai2021poso}.
\end{itemize}
\label{overwhelming}

Given the limitations of existing methods, we question \textbf{\emph{ the rationale of classical hard partitioning}}. Domain divisions within an app are often subjective, lacking comprehensive prior knowledge. User behavior can also be influenced by random factors—for example, a user might purchase late-night snacks via both the homepage and a sub-domain based on mood. Similarly, merchants often appear across multiple domains, reflecting the high user/item overlap (Table \ref{tab:overlap}). Thus, domain boundaries are far less rigid than traditionally assumed.
Based on the above observation, we explored \textbf{\emph{using dominant domain data to address domain imbalance}}. \textbf{S}oft-partitioned \textbf{S}emi-supervised \textbf{C}ollaborative \textbf{T}ransfer \textbf{L}earning (\textbf{SSCTL}), comprising two modules: the \textbf{I}nstance \textbf{S}oft-partitioned \textbf{C}ollaborative \textbf{T}raining (\textbf{ISCT}) process and the \textbf{S}oft-partitioned \textbf{D}omain \textbf{D}ifferentiation Network (\textbf{SDDN}), which addresses the overfitting issue within specific parameters and the overwhelming phenomenon within shared parameters, respectively. More specifically, ISCT treats dominant domain samples as unlabeled data, generating pseudo-labels with weights to enrich non-dominant domain data and mitigate overfitting. SDDN utilizes soft-partitioned domain information to generate dynamic parameters, reducing the overwhelming effect in shared parameters by shifting focus to non-dominant domain samples.

The main contributions can be summarized as follows: (1) We propose a novel soft-partitioning method that autonomously extracts domain information, contrasting the traditional hard-partitioning approach. (2) We introduce SSCTL, a multi-domain recommender combining ISCT and SDDN, which leverages dominant domain data to address overfitting in specific parameters and the overwhelming effect in shared parameters. (3) Extensive online and offline experiments validate the effectiveness of SSCTL.

\begin{table}\footnotesize
  \caption{The dominant domain is denoted as D1, while others are non-dominant domains.}
  \label{tab:overlap}
  \vspace{-0.8em}
  \renewcommand{\arraystretch}{0.75}
  \begin{tabular}{ccccccc}
    \toprule
        &   D1& D2& D3& D4& D5& D6 \\
    \midrule
    Items' Overlap& - & 82.61\% & 85.75\% & 74.93\% & 75.00\% & 74.53\% \\
    User' Overlap& - & 90.97\% & 91.96\% & 93.70\% & 99.12\% & 98.72\% \\
    \midrule
    Proportion & 81.16 \% & 12.57\% & 3.52\% & 1.14\% & 1.06\% & 0.59\% \\
  \bottomrule
\end{tabular}
\end{table}

\section{PRELIMINARIES}
We first formalize the multi-domain CTR prediction problem. Let \(\mathcal{X}\) denote the feature space (user, item, and context features) and \(\mathcal{Y}\) the label space (binary click indicators). For a platform with \(N\) domains \(\{D_1, D_2, ..., D_N\}\), the instance set of domain \(k\) is represented as \(\mathcal{D}_k=\{(x^k_i,y^k_i)\}^{|\mathcal{D}^k|}_{i=1} \), where:
    \(x^k_i \in \mathcal{X}\) is the feature of the i-th instance in domain \(k\),
    \(y^k_i \in \mathcal{Y}\) is the binary label of the i-th instance in domain \(k\).
The goal is to train a ranking model \(f_{\Theta}\), parameterized by \(\Theta\) to predict the label \(y^k_i\) accurately, and \(\hat{y}^k_i=f_{\Theta}(x^k_i)\).

\section{METHODOLOGY}

\begin{figure}[ht]
    \centering
    \includegraphics[width=1.0\linewidth]{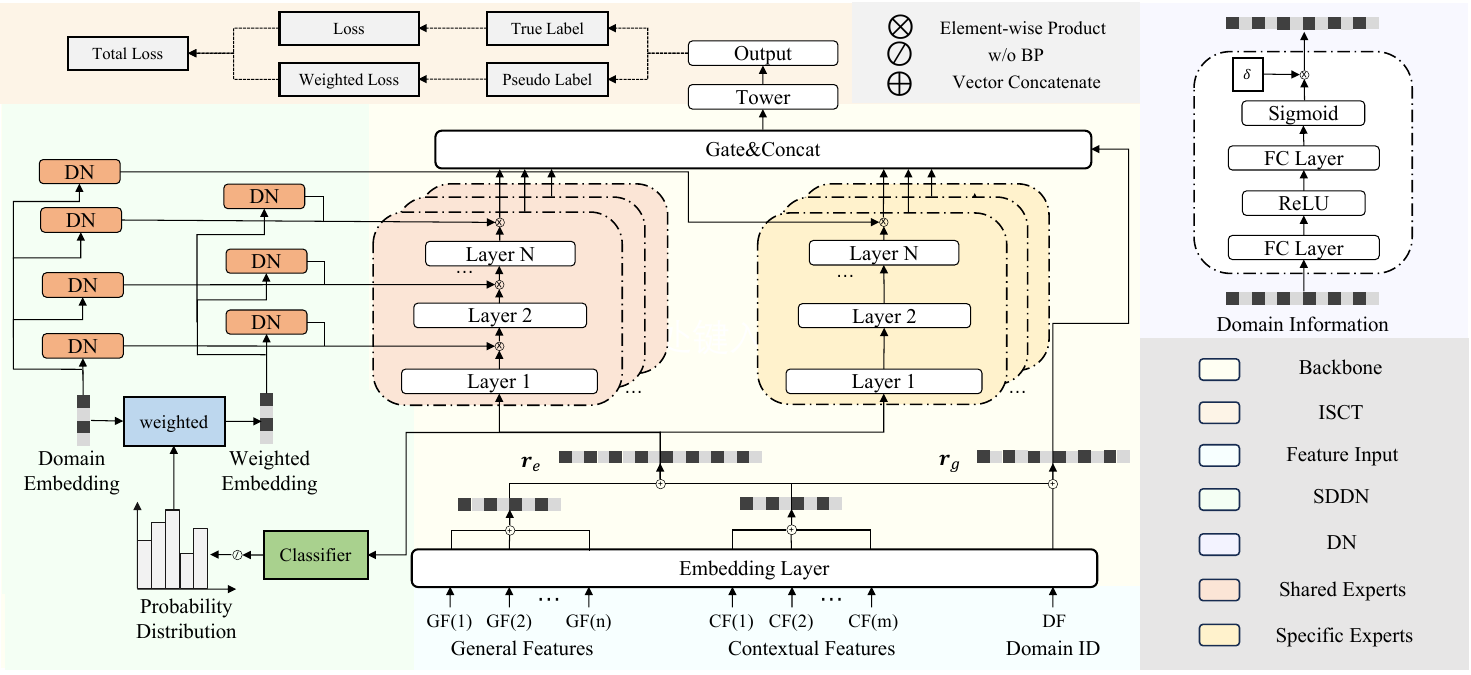}
    \caption{The Overall Framework of SSCTL. }
    \label{fig:SSCTL}
\end{figure}

\subsection{Backbone}
\label{sec: backbone}

Before introducing our model, we first outline the backbone framework. As shown in the yellow section of Figure \ref{fig:SSCTL}, we adopt a structure similar to CGC \cite{tang2020progressive}, consisting of five components: a feature embedding layer, domain-shared experts, domain-specific experts, a gating network, and a tower for prediction. Except for domain-specific experts, all other components are shared across domains. The feature embedding layer \(E (\cdot)\) transforms input features into embeddings as follows: $\textbf{r}_e =E(\mathcal{F}_G) \oplus E(\mathcal{F}_C),\ \textbf{r}_g =E(\mathcal{F}_C) \oplus E(\mathcal{F}_D),$
where \(\mathcal{F}_G\), \(\mathcal{F}_C\), \(\mathcal{F}_D\) are general, contextual, and domain features, respectively. Here, domain features correspond to the domain indicator. \(\textbf{r}_e\) is fed into experts, while \(\textbf{r}_g\) serves as input to the gating network. Unlike prior works \cite{li2020improving,STAR}, we do not feed domain features directly into experts; instead, these features are concatenated with contextual features (e.g., meal time) to guide network differentiation, as user behavior varies across scenarios. For clarity, we append a superscript \(k\) to \(\textbf{r}_e\) and \(\textbf{r}_g\) to indicate their association with the \(k\)-th domain \(\mathcal{D}_k\) (\(k=0,1,...,N-1\)).

As for experts, both the domain-shared experts and domain-specific experts consist of MLPS with \(L\) layers. There are $m$ domain-shared experts denoted as \(f_{i}(\cdot)\) (\(i = 1,...,m\)). Besides, each domain has a domain-specific expert, denoted as \(g^k(\cdot)\) (\(k=0,1,..., N-1\)).  The experts take embeddings \(\textbf{r}_e\) as input and generate hidden representations as follows:
\begin{equation}
    \textbf{h}^{k}_i = f_i(\textbf{r}^k_e),\quad
    \textbf{s}^k =g^k(\textbf{r}^k_e),
\end{equation}
\(\textbf{h}^{k}_i\) is the output of the \(i\)-th shared expert and \(\textbf{s}^k\) is the output of the domain-specific expert for \(\mathcal{D}_k\). A gating network is then employed to aggregate these representations. It generates weights \(\textbf{w}\) through a softmax function applied to the output of an MLP, and combines the representations as follows:
\begin{equation}
\begin{split}
     \textbf{w} = softmax(MLP(\textbf{r}_g)),\quad 
     \textbf{z} = \sum_{1 \leq i \leq m}w_i\textbf{h}^{k}_i \oplus \textbf{s}^k,
\end{split}
\end{equation}
where \(\textbf{z}\) represents the aggregated hidden representations. We ultimately input \(\textbf{z}\) into the tower \(t(\cdot)\) to obtain the result $\hat{y}=t(\textbf{z})$. \(\hat{y}\) denotes the final predicted user behavior label. 

\subsection{Instance Soft-partitioned Collaborative Training}
\label{ISCT}
In real-world e-commerce platforms, multiple recommendation domains coexist. Traditional hard-partitioning models train domain-specific parameters $g^k$ using data from a single domain. However, this approach has notable flaws: (1) Domain division and traffic allocation are predefined and lack robust consideration. (2) There is significant overlap of users/items across domains, with user purchase behavior often being random. (3) Data distribution is highly imbalanced, leading to overfitting for non-dominant domains. To address these issues, we propose ISCT. ISCT treats data from dominant domains as unlabeled and generates pseudo-labels for them, augmenting the data for non-dominant domains and mitigating overfitting of domain-specific parameters.

\begin{figure}
    \centering
    \setlength{\abovecaptionskip}{10pt}
    \includegraphics[width=1.0\linewidth]{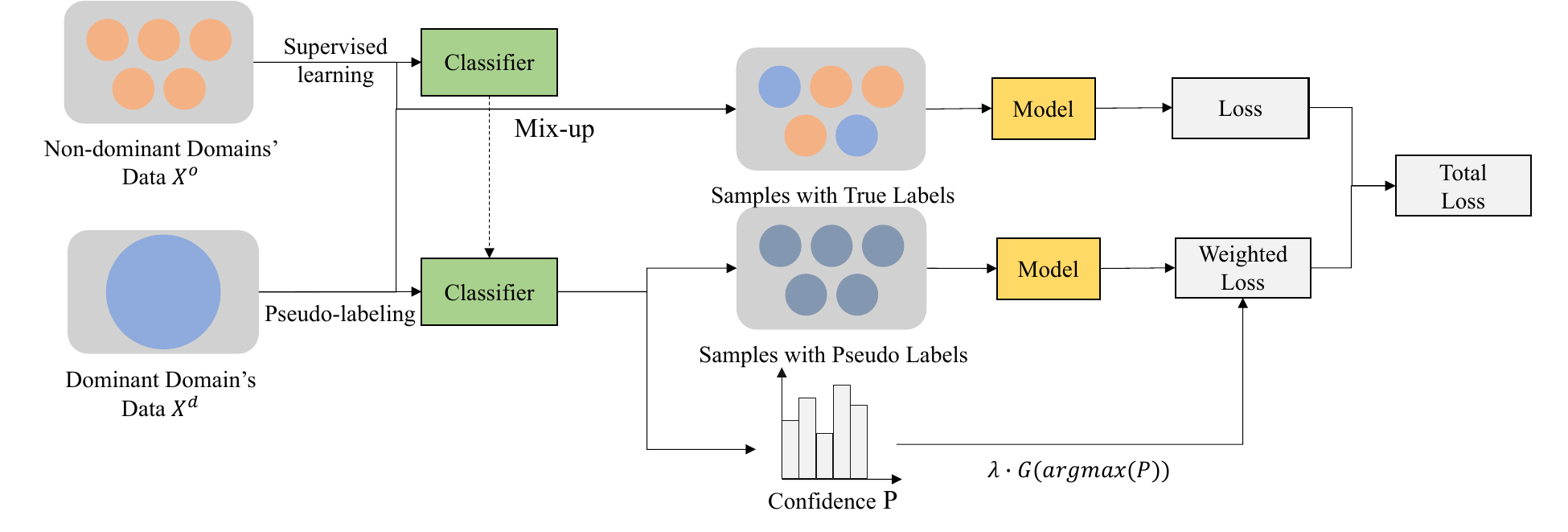}
    \caption{The Overall Procedure of ISCT. 
    }
    \label{fig:SCT}
\end{figure}

The ISCT process is as follows (illustrated in Figure \ref{fig:SCT}): 1) Partition data into subsets: dominant domain data \(\mathcal{X}^d\) and non-dominant domain's data \(\mathcal{X}^o\). 2) Train a classifier \(C(\cdot)\) on \(\mathcal{X}^o\)  using true labels from the \(N-1\) non-dominant domains. 3) Use \(C(\cdot)\) to generate pseudo-labels for \(\mathcal{X}^d\).

\begin{equation}
\begin{split}
    \textbf{p}^d_i=C(x^d_i),\quad  c_i=\max(\textbf{p}^d_i),\quad
    \overline{y}^d_i=k^*=\underset{j={1,2,..,N-1}}{\arg\max}p_{i,j}^d.
\end{split}
\end{equation}
\(\overline{y}^d_i\) is the pseudo-label for \(x^d_i\), and \(c_i\) is the confidence score. Samples with pseudo-labels form a new set \(\mathcal{X}^p\), which serves as auxiliary data for training. The original labels are retained, while the classifier \(C(\cdot)\) \textbf{is preserved for subsequent domain-related tasks}.

While pseudo-labels can be assigned to the dominant domain's data, low-confidence pseudo-labels often introduce noise. Traditional pseudo-labeling methods (e.g., FixMatch \cite{sohn2020fixmatch}) address this by applying fixed confidence thresholds to filter out low-confidence samples. However, we argue that higher-confidence samples are inherently less error-prone. To address this, we adopt the truncated Gaussian function \(G(\cdot)\) from SoftMatch \cite{chen2023softmatch} to compute sample weights based on pseudo-label confidence. The Gaussian distribution’s maximum entropy property ensures robust generalization, as demonstrated theoretically and empirically in SoftMatch.

This approach maximizes data utilization from the dominant domain while reducing the impact of noisy samples. The weight for each pseudo-label is calculated as:
\begin{equation}
    w_i=G(c_i)=
\begin{cases}
    \exp(-\frac{(c_i-\mu_t)^2}{2\sigma_t^2}),  & c_i\leq \mu_t \\
     1,   & otherwise.
\end{cases}
\end{equation}
\(\mu\) and \(\sigma\) are Gaussian parameters. Since the true parameters cannot be directly obtained, we use the Exponential Moving Average (EMA) during training to estimate them efficiently without added computational cost. This ensures higher weights for samples with higher pseudo-label confidence.

Finally, we combine samples with true labels and pseudo-labels into a single dataset for training. The total loss function is:
\begin{equation}
    \mathcal{L}_{total}=\sum_{x_i\in \mathcal{X}^d, \mathcal{X}^o}\mathcal{L}(x_i) +\lambda\sum_{x_j \in \mathcal{X}^p}w_j\mathcal{L}(x_j),
\end{equation}
where \(\lambda\) is a hyper-parameter controlling the contribution of pseudo-labels. This approach balances accuracy and noise reduction, improving model performance.

\subsection{Soft-partitioned Domain Differentiation Network}
In our business, data distribution imbalance across domains is severe, with one domain accounting for over 80\% of traffic. This causes the shared parameters to be dominated by the data from the dominant domain, limiting the model’s ability to leverage knowledge from other domains and resulting in sub-optimal performance. To address this, we propose the SDDN, shown in the green part of Figure \ref{fig:SSCTL}. SDDN enhances the influence of non-dominant domain samples on shared parameters by dynamically generating parameters based on domain information \cite{chang2023pepnet, M2M}. For example, domain indicators can be used as inputs for gating networks to produce scale vectors that adjust the intermediate layer outputs. This approach is similar to STAR \cite{STAR} but introduces dynamically generated parameters, effectively differentiating the main network into domain-specific sub-networks \cite{dai2021poso}.

Traditional multi-domain recommendation models often rely solely on domain indicators for hard partitioning. However, as discussed in Section \ref{ISCT}, our business involves complex rules (e.g., category, time-slot, and special business support rules) that hard partitioning fails to capture. To address this, SDDN adopts a soft-partitioning mechanism. Using the classifier \(C(\cdot)\) from ISCT, we generate a domain probability distribution for each sample, which effectively captures domain-related information at the sample level. This reflects both the commonalities and differences among domains under various rules, enabling more effective network differentiation compared to hard partitioning.

We input \(\textbf{r}_e\) into classifier \(C(\cdot)\) to obtain a probability distribution \(\textbf{p}\), which is then used to weight the domain embeddings \(\textbf{e}_d\). The weighted embedding \(\textbf{e}_w\) is calculated as:
\begin{equation}
\begin{split}
    \textbf{p} = C(\textbf{r}_e),  \quad  \textbf{e}_d = E(\mathcal{F}_d),\quad
    \textbf{e}_w = \sum_{j=1,2,...,N-1} p_j *  E(\mathcal{F}_d)_j,
\end{split}
\end{equation}
where \(p\) is an \(N-1\)-dimensional vector, and \(\textbf{e}_w\) is the weighted sum of embeddings for the \(N-1\) non-dominant domains. Both
\(\textbf{e}_d\) and \(\textbf{e}_w\) are fed into the \textbf{D}ifferentiation \textbf{N}etwork (\textbf{DN}). 
This mechanism integrates hard and soft partitioning, allowing the model to incorporate knowledge from non-dominant domains even when the true label belongs to the dominant domain. The output of \(C(\cdot)\) directly matches this structure, requiring no additional adjustments.
The DN consists of two fully connected layers: the first uses ReLU activation, and the second uses Sigmoid. Denoting input as \(\mathbf{x}\) and the fully connected layer as \(FC(\cdot)\),  the DN process is:
\begin{equation}
    DN(\textbf{x})=\delta * sigmoid(FC(ReLu(FC(\textbf{x})))).
\end{equation}
\(\delta\) is a hyperparameter to constrain the scale vector (set to 2 in our case) \cite{chang2023pepnet,yang2022adasparse}. Each shared layer has a corresponding DN, and for layer \(l\), the differentiation process is:
\begin{equation}
    \begin{split}
     \mathbf{\gamma}_{ld}&=DN_l(\mathbf{e}_d),\
     \quad \mathbf{\gamma}_{lw}=DN_l(\mathbf{e}_w),\\
     \mathbf{h_l^{\prime}}&=FC_l(h_{l-1}),\quad
     \mathbf{h_l} = \sqrt{\mathbf{\gamma}_{ld} * \mathbf{\gamma}_{lw}} \otimes \mathbf{h_l^{\prime}}.
    \end{split}
\end{equation}
\(\gamma_{ld}\) and \(\gamma_{lw}\) are scale vectors with dimensions matching the output of layer \(l\). \(\textbf{h}_l^{\prime}\) and \(\textbf{h}_l\) are the hidden and scaled hidden vectors, respectively, while \(\otimes\) denotes element-wise multiplication. The square root ensures consistent feature scaling. SDDN is plug-and-play and exhibits excellent scalability.


\subsection{Discussion}
We propose SSCTL to address challenges on our platform, but it can be easily adapted to other platforms and offers valuable insights for researchers.
(1) On our platform, the main homepage drives most traffic, a scenario common across e-commerce platforms, where homepages recommend all items and exhibit high user-item overlap across domains. This inspired our approach.
(2) 
While our method is motivated by this overlap, \textit{\textbf{it is not restricted to domains with high overlap}}. Instead, we assume that behavioral patterns from dominant domains can inform non-dominant ones, enabling effective transfer. Notably, \textit{\textbf{SSCTL does not rely on user or item IDs as features}}, yet it performs well, demonstrating its robustness beyond the high-overlap assumption.

\section{EXPERIMENTS}
In this section, we conduct extensive experiments both online and offline to evaluate the proposed method and answer the following questions.
\textbf{RQ1} \ How does the proposed method perform compared with state-of-the-art methods? \textbf{RQ2}  What factors affect the performance of multi-domain recommenders? 
\textbf{RQ3} \ How does SSCTL perform in real-world scenarios?

  

\subsection{Experimental Settings}
We evaluate our proposed method on two datasets: Ali-CCP\footnote{https://tianchi.aliyun.com/dataset/408} \cite{ma2018entire}, a public dataset, and MT-takeaway, an industrial dataset sampled from the Meituan Takeaway platform. The evaluation metrics include AUC (Area Under the Curve) and RImp (Relative Improvement) \cite{xi2021modeling,SAR-Net,wang2022causalint}.
To validate our model, we compared it with five categories of models: MLP, General\cite{guo2017deepfm,lian2018xdeepfm}, MTL (Multi-task Learning)\cite{ma2018modeling,tang2020progressive}, MDR (Multi-domain Recommendation)\cite{li2020improving,STAR,yang2022adasparse}, and MTMDR (Multi-task Multi-domain Recommendation)\cite{zhou2023hinet,chang2023pepnet}. Except for Single models trained on specific domains, all others were trained across all domains. For MTL models, domains were treated as separate tasks, while MTMDR models focused solely on CTR prediction.
Model settings included an embedding size of 10, hidden layers fixed to [256, 128, 64], Adam optimizer (learning rate = 1e-3), batch size = 4096, dropout = 0.2, and BatchNorm. For SSCTL, \(\lambda\) was set to 0.7. To address data imbalance in non-dominant domains, we applied Focal Loss \cite{lin2017focal} to the classifier. Domains \#1 and D1 were considered dominant.

\begin{table}\tiny
    \centering
    \setlength{\tabcolsep}{0.2pt}
    \renewcommand{\arraystretch}{0.6}
    \caption{Performance Comparison: The overall performance over Ali-CCP and MT dataset. The best and second-best results are highlighted in boldface and underlined respectively. $\star$ represents significance level $p$-value $<$ 0.05.}
    \vspace{-1.0em}
    \label{tab: performance}
    \begin{tabular}{ccccccccccccccc}
    \toprule

  \multirow{2}{*}{D} & \multirow{2}{*}{Metric} & \multicolumn{2}{c}{MLP} & \multicolumn{2}{c}{General} & \multicolumn{3}{c}{MTL} & \multicolumn{3}{c}{MDR} & \multicolumn{2}{c}{MTMDR} & \multirow{2}{*}{SSCTL} \\
  \cmidrule(lr){3-4} \cmidrule(lr){5-6} \cmidrule(lr){7-9} \cmidrule(lr){10-12} \cmidrule(lr){13-14}
                         &                         & Single & Mixed & DeepFM & xDeepFM & SBTM & MMoE & PLE & HMoE & STAR & AdaSparse & HiNet & PEPNet &  \\
  \midrule

\multicolumn{15}{c}{\textbf{Ali-CCP}}\\
    
    \midrule
        \multirow{2}{*}{\#1}&AUC&0.6276&0.6309&0.6316&0.6325&0.6305&0.6302&0.6299&0.6307&0.6309&\underline{0.6310}&0.6298&0.6304&{$\mathbf{0.6331}^\star$}\\
                                                    &RImp&-&+0.53\%&+0.65\%&+0.79\%&+0.47\%&+0.41\%&+0.37\%&+0.50\%&+0.53\%&+0.55\%&+0.35\%&+0.45\%&+0.86\%\\
        \hline 
                                \multirow{2}{*}{\#2}&AUC&0.6235&0.6271&0.6276&0.6284&0.6243&0.6252&0.6250&0.6257&\underline{0.6278}&0.6268&0.6250&0.6275&$\mathbf{0.6301}^\star$\\
                                                    &RImp&-&+0.56\%&+0.65\%&+0.78\%&+0.12\%&+0.27\%&+0.25\%&+0.35\%&+0.68\%&+0.52\%&+0.24\%&+0.64\%&+1.17\%\\
                                                    
        \hline 
                                \multirow{2}{*}{\#3}&AUC&0.5530&0.6022&0.6013&0.6038&0.5853&0.5997&\underline{0.6066}&0.5760&0.5773&0.6022&0.5990&0.6024&$\mathbf{0.6095}^\star$\\
                                                    &RImp&-&+8.95\%&+8.75\%&+9.21\%&+5.86\%&+8.45\%&+9.70\%&+4.17\%&+4.41\%&+8.91\%&+8.33\%&+8.95\%&+10.23\%\\
        \midrule

        \multicolumn{15}{c}{\textbf{MT}}\\
        
        \midrule
        \multirow{2}{*}{D1}&AUC&0.6925&0.6925&0.6938&0.6946&0.6929&0.6935&0.6942&0.6935&0.6938&0.6939&0.6946&\underline{0.6947}&\textbf{0.6951}\\
                                                &RImp&-&+0.00\%&+0.18\%&+0.29\%&+0.06\%&+0.14\%&+0.25\%&+0.12\%&0.18\%&+0.20\%&+0.30\%&+0.32\%&+0.38\%\\
        \hline 
                                \multirow{2}{*}{D2}&AUC&0.6501&0.6725&0.6753&0.6758&0.6690&0.6747&0.6754&0.6684&0.6718&\underline{0.6759}&\underline{0.6759}&0.6741&$\textbf{0.6783}^\star$\\
                                                &RImp&-&+3.44\%&+3.88\%&+3.95\%&+2.90\%&+3.78\%&+3.90\%&+2.82\%&+3.22\%&+3.96\%&+3.96\%&+3.70\%&+4.35\%\\
        \hline 
                                \multirow{2}{*}{D3}&AUC&0.7111&0.7219&0.7216&0.7225&0.7199&0.7226&0.7224&0.7218&0.7220&\underline{0.7228}&0.7220&0.7214&$\textbf{0.7245}^\star$\\
                                                &RImp&-&+1.52\%&+1.48\%&+1.60\%&+1.24\%&+1.62\%&+1.58\%&+1.51\%&+1.53\%&+1.64\%&+1.53\%&+1.45\%&+1.89\%\\
        \hline 
                                \multirow{2}{*}{D4}&AUC&0.6560&0.6853&0.6844&0.6858&0.6772&0.6855&0.6835&0.6814&0.6819&0.6852&\underline{0.6865}&0.6853&$\textbf{0.6907}^\star$\\
                                                &RImp&-&+4.48\%&+4.36\%&+4.56\%&+3.25\%&+4.51\%&+4.22\%&+3.90\%&+3.97\%&+4.47\%&+4.68\%&+4.49\%&+5.31\%\\
        \hline 
                                \multirow{2}{*}{D5}&AUC&0.6207&0.6510&0.6529&0.6541&0.6445&0.6552&\underline{0.6553}&0.6514&0.6509&0.6515&0.6552&0.6532&$\textbf{0.6591}^\star$\\
                                                &RImp&-&+4.87\%&+5.18\%&+5.38\%&+3.82\%&+5.55\%&+5.56\%&+4.94\%&+4.86\%&+4.95\%&+5.56\%&+5.22\%&+6.19\%\\
        \hline 
                                \multirow{2}{*}{D6}&AUC&0.6061&0.6700&0.6673&0.6722&0.6613&0.6740&0.6691&0.6657&0.6669&0.6753&0.6734&\underline{0.6777}&$\textbf{0.6817}^\star$\\
                                                &RImp&-&+10.53\%&+10.10\%&10.90\%&+9.10\%&+11.19\%&+10.39\%&+9.82\%&+10.02\%&+11.40\%&+11.10\%&+11.48\%&+12.47\%\\
        \bottomrule
    \end{tabular}
    \vspace{-1.8em}
\end{table}

\subsection{Performance Comparison(RQ1\&RQ2)}
%
Table \ref{tab: performance} presents the results on two datasets across multiple domains. Notably, a 0.002 AUC improvement in the dominant domain is significant due to the abundance of data. Key observations include:
\begin{itemize}[leftmargin=*]
    \item \textbf{SSCTL outperforms all baselines.} SSCTL consistently achieves the best performance across all datasets and domains, demonstrating its effectiveness in multi-domain recommendations.

    \item \textbf{Sparsity in non-dominant domains limits shared-specific models.} Models like SharedBottom, HMoE, and STAR perform poorly on sparse non-dominant domains (e.g., \#3, D5, D6), as sparse data causes overfitting in specific parameters. In contrast, SSCTL leverages dominant domain data to augment non-dominant domains, significantly improving performance.
    \item \textbf{Dynamic parameters enhance non-dominant domain performance.} Models like AdaSparse, PEPNet, and SSCTL, which use dynamic parameters for network differentiation, outperform others on non-dominant domains. This approach mitigates the overwhelming effect by amplifying the impact of non-dominant domain samples on shared parameters.

\end{itemize}

\begin{table} \small
    \centering
    \renewcommand{\arraystretch}{0.75}
    \caption{Result of Online A/B Test}
    \vspace{-1.0em}
    \begin{tabular}{cccccc}
    \toprule
         &  1 & 2 & 3 & 4 & 5\\
    \midrule
       PVCTR & +0.57\% & +0.80\% & +0.22\% & +1.40\% &+1.69\% \\
    \midrule
       GMV & +0.54\% & +1.17\% & +1.63\% &+2.06\%&+2.90\%\\
    \bottomrule
    \end{tabular}
    \vspace{-1.5em}
    \label{Online}
\end{table}

\subsection{Online Experiments (RQ3)}
We deployed SSCTL across five domains on the Meituan Takeaway platform and conducted a 10-day online A/B test. As shown in Table \ref{Online}, SSCTL consistently outperformed the baseline model, improving GMV (Gross Merchandise Volume) by 0.54\%–2.90\% and CTR (Click-Through Rate) by 0.22\%–1.69\% across all domains. These results demonstrate SSCTL’s effectiveness in real-world applications.

\section{CONCLUSION}
This work identifies two key challenges in shared-specific architectures caused by imbalanced data distribution in multi-domain recommendations and the limitations of commonly used hard-partitioning methods. To address these, we propose SSCTL, which integrates ISCT and SDDN to mitigate overfitting in specific parameters and reduce the overload in shared parameters. Extensive offline and online experiments validate the effectiveness of SSCTL. We hope this study offers insights into leveraging dominant domain data in MDR problems.

\section*{Acknowledgments}
The research work is supported by the National Key Research and Development Program of China under Grant Nos. 2024YFF0729003, the National Natural Science Foundation of China under Grant NO. 62176014, the Fundamental Research Funds for the Central Universities.

\section*{GenAI Usage Disclosure}
We employed AI exclusively for grammatical refinement and sentence polishing.

\bibliographystyle{ACM-Reference-Format}
\balance
\bibliography{reference}

\end{document}